\documentclass[structabstract]{aa}  
\usepackage{graphicx} 
\usepackage{txfonts}
\usepackage{natbib}
\usepackage{breqn}
\usepackage{lscape}
\usepackage{hyperref}

\newcommand{\HII}{\mbox{H\,{\sc ii}}}
\newcommand{\HeI}[1]{\mbox{He\,{\sc i}~$\lambda${#1}}}

\newcommand{\CII}[1]{\mbox{C\,{\sc ii}~$\lambda${#1}}}
\newcommand{\CIII}[1]{\mbox{C\,{\sc iii}~$\lambda${#1}}}

\newcommand{\SiIId}[1]{\mbox{Si\,{\sc ii}~$\lambda\lambda${#1}}}
\newcommand{\SiIII}[1]{\mbox{Si\,{\sc iii}~$\lambda${#1}}}
\newcommand{\SiIIIt}[1]{\mbox{Si\,{\sc iii}~$\lambda\lambda\lambda${#1}}}

\newcommand{\MgII}[1]{\mbox{Mg\,{\sc ii}~$\lambda${#1}}}

\newcommand{\mci}[1]{\multicolumn{1}{c}{#1}}

\begin{document}

   \title{Spatially resolved spectroscopy of close massive visual binaries \linebreak
          with HST/STIS: I. Seven O-type systems.}
   \titlerunning{HST/STIS spectroscopy of close O-type visual binaries.}

   \author{J. Ma\'{\i}z Apell\'aniz \inst{1}
           \and
           R. H. Barb\'a\inst{2}
           }

   \institute{Centro de Astrobiolog\'{\i}a, CSIC-INTA. Campus ESAC. 
              C. bajo del castillo s/n. 
              E-28\,692 Villanueva de la Ca\~nada, Madrid, Spain.\linebreak
              \email{jmaiz@cab.inta-csic.es} \\
              \and
              Departamento de Astronom{\'\i}a, Universidad de La Serena.
              Av. Cisternas 1200 Norte.
              La Serena, Chile. \\
             }

   \date{Received 14 Feb 2020 / Accepted 27 Feb 2020}

 
  \abstract
   {Many O-type stars have nearby companions whose presence hamper their characterization through spectroscopy.}
   {We want to obtain spatially resolved spectroscopy of close O-type visual binaries to derive their spectral types.}
   {We use the Space Telescope Imaging Spectrograph (STIS) of the \textit{Hubble Space Telescope} (HST) to obtain long-slit blue-violet spectroscopy
    of eight Galactic O-type stars with nearby visual companions and use spatial-profile fitting to extract the separate spectra. We also use the ground-based
    Galactic O-Star Spectroscopic Survey (GOSSS) to study more distant visual components.}
   {We spatially resolve seven of the eight systems, present spectra for their components, and obtain their spectral types. Those seven multiple systems are
    $\iota$~Ori~Aa,Ab,B, 15~Mon~Aa,Ab,C, $\tau$~CMa~Aa,Ab,B,C,D,E, HD~206\,267~Aa,Ab,C,D, HD~193\,443~A,B, HD~16\,429~Aa,Ab, and IU~Aur~A,B. This is the first time that spatially
    resolved spectroscopy of the close visual binaries of those systems is obtained. We establish the applicability of the
    technique as a function of the separation and magnitude difference of the binary.}
   {}

   \maketitle
%

\section{Introduction}

$\,\!$\indent For many years now we have been studying the multiplicity of O stars using different spectroscopic, imaging, and interferometric techniques. Such a 
combination is needed because even though most (if not all) massive stars are born in multiple systems, the different periods, inclinations, mass ratios, distances, and
extinctions make each detection technique more adequate for one or another of the hundreds of O-type multiple systems known. One preliminary aspect of the study of such
systems is their identification as O stars, something we are doing with the Galactic O-Star Spectroscopic Survey (GOSSS, \citealt{Maizetal11}). The subsequent cataloguing 
and spectroscopic characterization of multiple O-type systems is being done with the MONOS project in the northern hemisphere (\citealt{Maizetal19b}, from now on MONOS-I) and 
with the OWN project in the southern hemisphere \citep{Barbetal10,Barbetal17}.

Massive multiple systems with small plane-of-the sky separations $d$ 
are typically studied using multi-epoch high-resolution spectroscopy, as most such systems known in the Galaxy have short
periods ($\lesssim$1~a) that yield significant differences in velocity between the two components that allow spectroscopic orbits to be calculated (possibly with the
help of eclipses for the shortest periods with high orbital inclinations). Multiple systems with $d$ of a few arcseconds or more can be easily resolved from the
ground and spectra for the different components thus obtained. This leaves a problematic $d$ range between $\approx$30~mas and $\approx$3\arcsec, where velocity
differences are usually small and the spectra cannot be easily separated from the ground\footnote{As discussed later on, those ranges are approximate and depend on the
magnitude difference between components.}. Furthermore, a large fraction of the O-type multiple systems are not binaries but triples, quadruples, or worse 
\citep{Sotaetal14,Maizetal19b} and in those cases where three or more massive stars are located within 3\arcsec, the identification of which star corresponds to which 
spectral type can become a difficult task. 

To tackle that problematic range, three years ago we developed the new lucky spectroscopy technique \citep{Maizetal18a}, which is the spectroscopic equivalent to lucky 
imaging \citep{Lawetal06,Baldetal08,Smitetal09}. With lucky spectroscopy one obtains a large number of short spectroscopic exposures, selects the ones with better seeing
qualities, combines the result, and extracts the resulting spectra fitting a double or triple profile. In \citet{Maizetal18a} we used the ISIS spectrograph at the 
William Herschel Telescope (WHT) at La Palma to spatially separate five close visual systems with massive stars and in subsequent observations we have been successful with 
several tens more (Ma{\'\i}z Apell\'aniz et al. in preparation). However, in the process it has 
also become clear that lucky spectroscopy has its limitations and that (at least in the configuration we are using) it cannot resolve systems in the lower part of the
30~mas-3\arcsec\ interval. If we want to obtain spatially separated spectra in the lower part of that range we need to resort to the {\it Hubble Space Telescope} (HST) and 
that is what we have done in this paper. In the next section we describe our data and methods, in section 3 we present our results, and in the last section we analyze them
and discuss our future work.

\begin{table*}
\caption{HST/STIS sample in this paper plus a system previously observed with this technique (HD~93\,129~Aa,Ab). The position angle $\theta$ corresponds to the 
angle of the slit as measured from the first to the second component while $d^\prime$ corresponds to the plane-of-the-sky separation between the two components
along the slit. An asterisk is added for those cases where $\Delta B$ or $d^\prime$ were fixed from the literature instead of measured from the data. The 
separation marked with a $\dagger$ was measured with the slit positioned far from the predicted position angle, making $d$ and $d^\prime$ significantly different 
(see text).  Measured uncertainties are typically 0.01~mag in $\Delta B$ and 1~mas in $d^\prime$.}
\label{sample}
\begin{center}
\begin{tabular}{lrrrccc}
\hline
name               & \mci{$\Delta B$} & \mci{$d^\prime$} & \mci{$\theta$} & WDS ID       & obs. date & res.? \\
                   & \mci{(mag)}      & (mas)            & (deg)          &              & (YYMMDD)  &       \\
\hline
$\zeta$~Ori~Aa,Ab  & 2.20*            &  39*             & 280            & 05407$-$0157 & 191118    & N     \\
$\iota$~Ori~Aa,Ab  & 3.16\phantom{*}  & 155*             &  98            & 05354$-$0555 & 191119    & Y     \\
15~Mon~Aa,Ab       & 1.55\phantom{*}  & 143\phantom{*}   & 269            & 06410$+$0954 & 191129    & Y     \\
$\tau$~CMa~Aa,Ab   & 0.90\phantom{*}  &  95\phantom{*}   & 310            & 07187$-$2457 & 200102    & Y     \\
$\tau$~CMa~Aa,E    & 4.47\phantom{*}  & 685$\dagger$     & 310            & 07187$-$2457 & 200102    & Y     \\
HD~206\,267~Aa,Ab  & 0.95\phantom{*}  &  64\phantom{*}   & 147            & 21390$+$5729 & 191206    & Y     \\ 
HD~193\,443~A,B    & 0.26\phantom{*}  & 138\phantom{*}   & 250            & 20189$+$3817 & 191013    & Y     \\
HD~16\,429~Aa,Ab   & 2.02\phantom{*}  & 275\phantom{*}   &  91            & 02407$+$6117 & 191126    & Y     \\
IU~Aur~A,B         & 2.04\phantom{*}  & 138*             & 232            & 05279$+$3447 & 191003    & Y     \\
HD~93\,129~Aa,Ab   & 1.23\phantom{*}  &  36*             &  14            & 10440$-$5933 & 100407    & Y     \\
\hline
\end{tabular}
\end{center}
\end{table*}

\section{Methods and data}

\subsection{Sample selection}

$\,\!$\indent To select our sample we used the Galactic O-Star Catalog (GOSC, \citealt{Maizetal04b,Maizetal17c}) and searched for bright Galactic O-type close
visual systems of astrophysical interest. More specifically, we searched for systems located in the separation-$B$ magnitude difference ($d-\Delta B$) plane not 
accessible to being spatially resolved using current lucky spectroscopy capabilities \citep{Maizetal18a} but that could be resolved in the blue-violet region using 
HST/STIS based on our previous past experience \citep{Maizetal17a}. Our full sample has ten such systems and is being observed with HST GO program 15\,815.
One of the systems ($\theta^1$~Ori~Ca,Cb) has not been observed as of the time of this writing and the visit for another one (HD~193\,322~Aa,Ab) failed to
acquire data correctly and will have to be repeated. Therefore, the sample for this paper includes only eight systems; the other two will be analyzed in
future papers. The systems are listed in Table~\ref{sample} where, for completeness, we have added another system (HD~93\,129~Aa,Ab) previously observed by 
us using the same technique with data from HST GO program 11\,784. 

\subsection{Data acquisition and processing}

$\,\!$\indent We used the same observing technique applied to the HD~93\,129~Aa,Ab case mentioned above \citep{Maizetal17a}.
We obtain STIS exposures with five G430M settings that yield a 
continuous coverage of the 3793-5104~\AA\ range\footnote{For HD~93\,129~Aa,Ab we covered a smaller region in the blue-violet region but we added H$\alpha$.}. To increase 
the effective spatial resolution a two-point dithering pattern along the slit is applied so that for each grating setting we end up with a 1024$\times$2048 
(spectral $\times$ spatial) pixel array. The wavelength coverage is determined by the number of grating settings that can be fit in an orbit (five), as our targets are bright 
and require only short exposures and most of the time in the orbit is actually consumed by overheads.

The position angle of the slit $\theta$ was chosen to be as close as the telescope scheduling allowed to the expected position angle between the two components 
(with a possible 180$^{\rm o}$ offset). We were able to achieve the desired value within 1$^{\rm o}$ in all cases except in two: $\zeta$~Ori~Aa,Ab, where the difference was
4$^{\rm o}$ and $\iota$~Ori~Aa,Ab, where the difference was 6$^{\rm o}$. Note that the cosines of those angles are larger than 0.99 so the difference between $d$ and its 
projection along the slit $d^\prime$ should be small. For $\tau$~CMa we unintentionally also recorded a third component in the slit (E). In that case, the Aa,E position angle is 
expected to be 267$^{\rm o}$ (there is a 180$^{\rm o}$ offset in some of the previously published data, see below), so there is a difference of 43$^{\rm o}$ with $\theta$. 
We use the 52X2 slit ($52\arcsec\times2\arcsec$ or $1024\times 39$ STIS/CCD pixels), so the expected 940~mas separation between Aa and E translates into 687~mas along the
slit ($y$ in the detector) and 641~mas in the direction perpendicular to it ($x$ in the detector), which corresponds well to the measured values (see below).
That may translate into a small slit loss not accounted for in our $\Delta B$ and into an artificial velocity shift for E (which we indeed detect).

\begin{figure*}
\centerline{\includegraphics[width=0.49\linewidth]{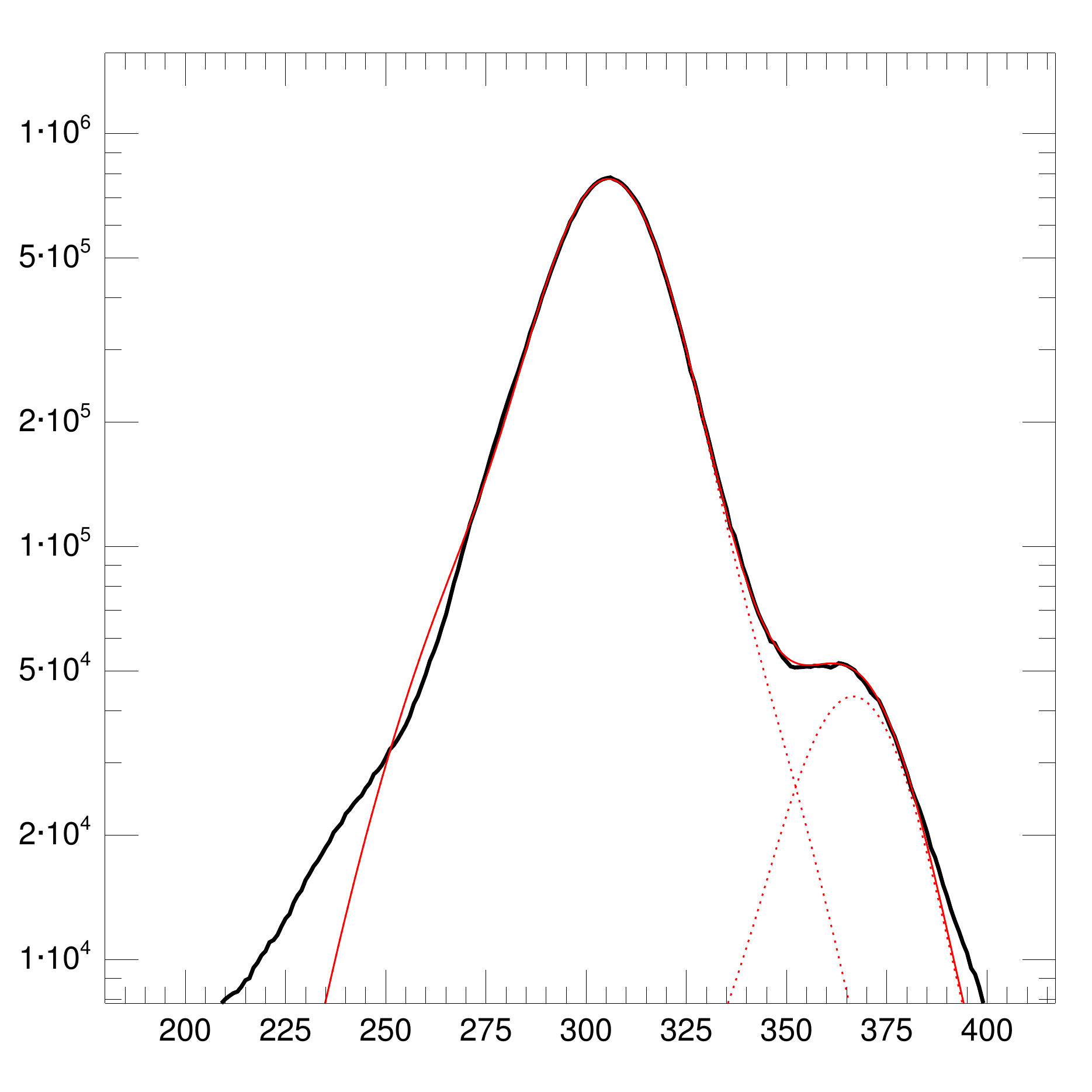} \
            \includegraphics[width=0.49\linewidth]{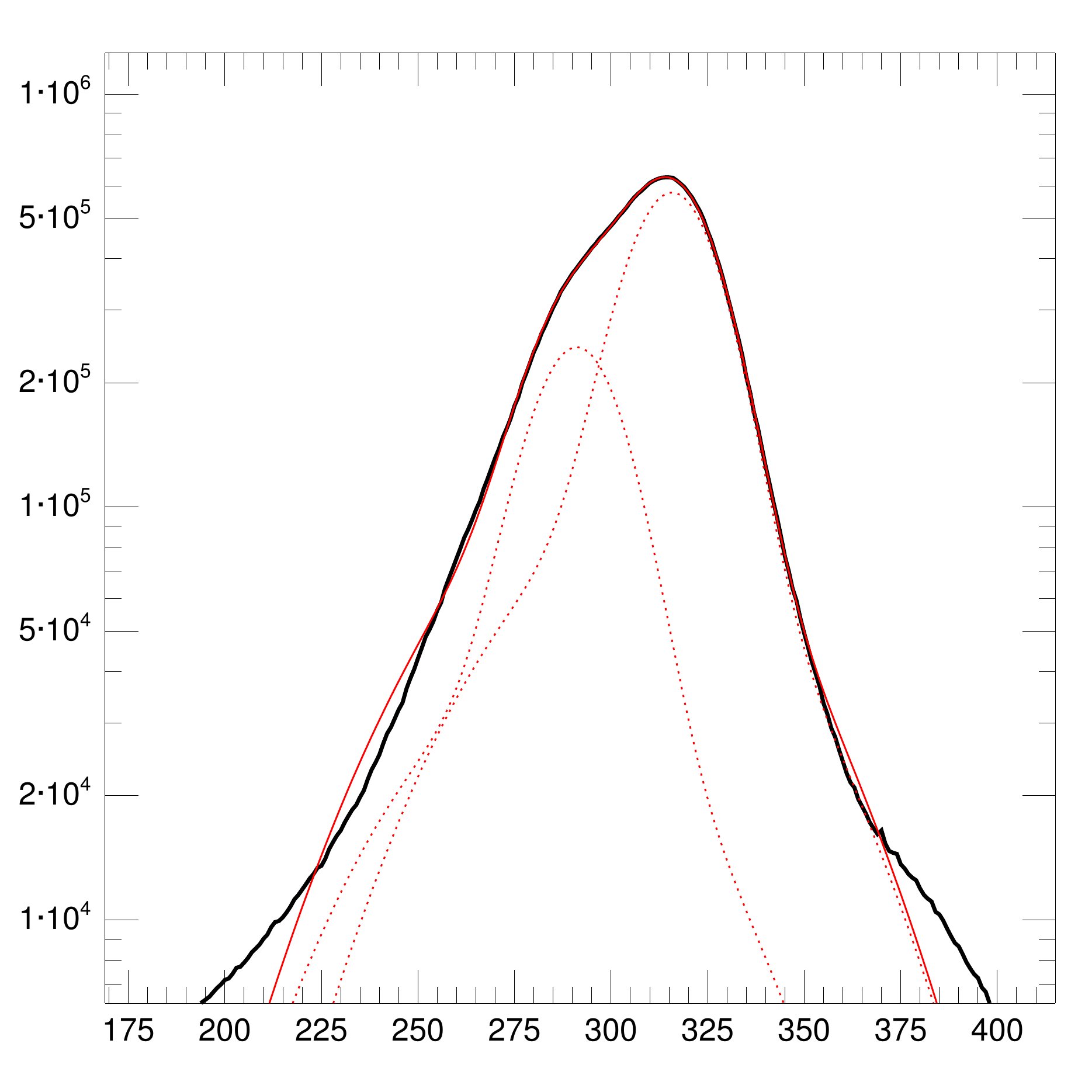}}
\caption{Spatially dithered spectral data collapsed along the trace used to fit the double spatial profile for two of the systems in this paper ($\iota$~Ori~Aa,Ab, left,
         and HD~206\,267~Aa,Ab, right, in both cases for the 3936 grating setting). The horizontal axes show the 20x dithered spatial pixels (1~dithered pixel~=~2.5~mas) and the 
         vertical axes the counts (in log scale). In each plot the black line is the data, the solid red line the double (Aa+Ab) profile, and the dotted red lines the individual 
         (Aa and Ab) profiles. The left plot is an example with relatively large separation and $\Delta B$ while the right one is an example of the opposite circumstances. See Fig.~1 
         of \citet{Maizetal18a} for equivalent plots using lucky spectroscopy.}
\label{spatialprofiles}
\end{figure*}

The spectra were extracted using the techniques derived for MULTISPEC \citep{Maiz05a}, which were based in part on the theoretical fundaments presented by 
\citet{Portetal04b} and that were further developed for the analysis of the STIS HD~93\,129~Aa,Ab data in \citet{Maizetal17a} and for lucky spectroscopy in \citet{Maizetal18a}. 
In the STIS CCD spectra the trace is nearly aligned with the $x$ axis of the detector but not exactly so. To determine the spatial profile that will be used to separate the two
components in each system, we first calculated the centroid in the $y$ direction at each of the 1024 $x$ positions of the 1024$\times$2048 pixel array for each grating setting 
independently. The resulting $y_c(x)$ was fitted with a third- or (if needed) a fourth-degree polynomial to determine the trace. As there is a large number of photons collected in each
exposure we can accurately determine a dithered spatial profile \citep{FrucHook02} and fit a functional form to it in the following manner:

\begin{itemize}
 \item We multiply $y_c(x)$ by ten, round off the result, and classify each $x$ position by its last digit (from 0 to 9), a value we will call $R$. This determines how is the trace 
       centered with respect to a pixel center at each value of $x$. Following the convention where a pixel center has an integer coordinate value and an edge a half-integer one, if 
       $R(x)$ is 0 then the trace there is centered at the pixel center and if it is 5 then the trace there is centered at the pixel edge. 
 \item Given the size of the pixel array and the smooth nature of the trace, a given value of $R$ (from 0 to 9) takes place at $\sim$102 $x$ positions. For each value of $R$
       we sum the data in all the corresponding $x$ positions (shifting them to adjust the position of the center) and normalize the result to obtain the spatial profile as a function of
       $R$. Note that since the $\sim$102 $x$ positions are scattered over the trace and the result is normalized, the result contains only spatial information, as all the spectral 
       information is effectively erased. This should not be a problem in principle, as the procedure is repeated for each grating setting and the wavelength range covered by each 
       setting is small enough for the spatial profile not to significantly depend on wavelength. 
 \item We now generate a dithered spatial profile by interspersing the profiles for the ten values of $R$. This profile has a pixel scale of 50~mas/20~=~2.5~mas, as the STIS CCD pixel
       size is 50~mas and we have dithered the result in two steps: first physically by placing the detector in two positions separated by 25~mas and then generating the profile for the
       ten values of $R$. 
 \item Finally, the result is fitted with a double functional form along with the magnitude difference $\Delta B$ and, possibly, $d$ (see Fig.~\ref{spatialprofiles} for two examples.)
       The separation is fitted or fixed from the literature depending on the position of the system in the $d$-$\Delta B$ plane: objects close to the boundary that delimits the 
       validity of the technique (Fig.~\ref{sep_DB}) cannot be fitted as the process does not uniformly converge to a solution consistent with literature values. The spatial profile is 
       fitted with a slight asymmetry to take into account charge transfer inefficiency (CTI) effects but we note the targets were placed at the E1 slit position (close to the amplifier 
       readout) to minimize such effects. 
\end{itemize}

Once a trace and a functional form are determined, the two spectra are fitted one $x$ value (i.e. one wavelength) at a time to extract the spectrum. Since the purpose of 
this paper is to obtain spectral classifications, the resulting spectra are rectified, degraded to a spectral resolution of $R = 2500$, and shifted to the reference frame of
each component, so that they can be compared with the GOSSS standards \citep{Maizetal11,Maizetal17c} using MGB \citep{Maizetal12,Maizetal15b}. 
MGB is an interactive code that overplots the spectrum with a library of standards, allowing the user to select the spectral type and luminosity class, artificially
broaden the standard to simulate the effect of the different rotation indices, and (if necessary) combine two standards with arbitrary $\Delta m$ and velocities to 
classify SB2 systems. In future works we will use the original data to study the extinction law and measure the velocities of the components.

\subsection{Additional data}

$\,\!$\indent Some of the systems in this paper have additional components at separations that allow for resolved spectroscopy from the ground. We have observed some of them
with GOSSS and presented their spectra and spectral classifications in previous papers (e.g. MONOS-I). Here we do the same for seven components that have not previously
appeared in GOSSS papers. The reader is referred to the GOSSS papers for details on the data acquisition and processing. Additionally, we have also used current and historical information
about component separations, position angles, and magnitude differences from the Washington Double Star Catalog (WDS, \citealt{Masoetal01}).

\section{Results}

$\,\!$\indent In this section we present our spectral classifications for the seven systems we have spatially resolved with STIS/HST. In each case we first describe
the multiplicity of the system. The spectral classifications are listed in Table~\ref{spclas} and the spectra are shown in Figure~\ref{spectra}. GOS/GBS/GAS stands for Galactic
O/B/A Star following the nomenclature in the GOSSS papers. No results are given for the eighth system, $\zeta$~Ori~Aa,Ab, as we were unable to spatially resolve the two 
components. GOSSS spectra and classifications for additional components of some of the multiple systems are given in Figure~\ref{spectraGOSSS} and Table~\ref{spclasGOSSS}, 
respectively.

\begin{table*}
\caption{Spectral classifications derived from HST/STIS data.}
\label{spclas}
\begin{center}
\begin{tabular}{lcllll}
\hline
name           & GOS/GBS ID             & \mci{spec.} & \mci{lum.}  & \mci{qual.} & \mci{sec.} \\
               &                        & \mci{type}  & \mci{class} &             &            \\
\hline
$\iota$~Ori~Aa & GOS 209.52$-$19.58\_01 & O8.5        & III         & \ldots      & sec        \\
$\iota$~Ori~Ab & GBS 209.52$-$19.58\_02 & B2:         & IV:         & \ldots      & \ldots     \\
15~Mon~Aa      & GOS 202.94$+$02.20\_01 & O7          & V           & ((f))z      & \ldots     \\
15~Mon~Ab      & GBS 202.94$+$02.20\_03 & B1:         & V           & n           & \ldots     \\
$\tau$~CMa~Aa  & GOS 238.18$-$05.54\_01 & O9.2        & Iab:        & \ldots      & \ldots     \\
$\tau$~CMa~Ab  & GOS 238.18$-$05.54\_02 & O9          & III         & \ldots      & \ldots     \\
$\tau$~CMa~E   & GOS 238.18$-$05.54\_03 & B2:         & V           & \ldots      & \ldots     \\
HD~206\,267~Aa & GOS 099.29$+$03.74\_01 & O5          & V           & ((fc))      & B0: V      \\
HD~206\,267~Ab & GOS 099.29$+$03.74\_03 & O9          & V           & \ldots      & \ldots     \\
HD~193\,443~A  & GOS 076.15$+$01.28\_01 & O8.5        & III         & ((f))       & \ldots     \\
HD~193\,443~B  & GOS 076.15$+$01.28\_02 & O9.2        & IV          & \ldots      & \ldots     \\
HD~16\,429~Aa  & GOS 135.68$+$01.15\_01 & O9.2        & III         & \ldots      & \ldots     \\
HD~16\,429~Ab  & GOS 135.68$+$01.15\_02 & O9.5        & IV          & \ldots      & \ldots     \\
IU~Aur~A       & GOS 173.05$-$00.03\_01 & O8          & IV          & (n)         & O9.5 IV(n) \\
IU~Aur~B       & GBS 173.05$-$00.03\_02 & B1:         & V           & \ldots      & \ldots     \\
\hline
\end{tabular}
\end{center}
\end{table*}

\begin{table*}
\caption{Spectral classifications derived from GOSSS data.}
\label{spclasGOSSS}
\begin{center}
\begin{tabular}{lcccllll}
\hline
name           & GBS/GAS ID             & R.A.         & dec.           & \mci{spec.} & \mci{lum.}  & \mci{qual.} \\
               &                        & (J2000)      & (J2000)        & \mci{type}  & \mci{class} &             \\
\hline
$\iota$~Ori~B  & GBS 209.53$-$19.58\_01 & 05:35:26.455 & $-$05:54:44.46 & B2          & V           & He poor     \\
15 Mon C       & GAS 202.93$+$02.20\_01 & 06:40:58.938 & $+$09:54:00.85 & A3          & V           & \ldots      \\
$\tau$ CMa B   & GBS 238.18$-$05.54\_04 & 07:18:43.098 & $-$24:57:15.80 & B2          & V           & n           \\
$\tau$ CMa C   & GBS 238.18$-$05.54\_05 & 07:18:43.537 & $-$24:57:12.96 & B5          & V           & nnn         \\
$\tau$ CMa D   & GBS 238.19$-$05.52\_04 & 07:18:48.532 & $-$24:56:55.97 & B0.7        & V           & \ldots      \\
HD 206\,267 C  & GBS 099.29$+$03.74\_05 & 21:38:58.887 & $+$57:29:14.64 & B0          & V           & (n)         \\
HD 206\,267 D  & GBS 099.29$+$03.74\_02 & 21:38:56.736 & $+$57:29:39.25 & B0.2        & V           & \ldots      \\
\hline
\end{tabular}
\end{center}
\end{table*}

\begin{figure*}
\centerline{\includegraphics[width=\linewidth]{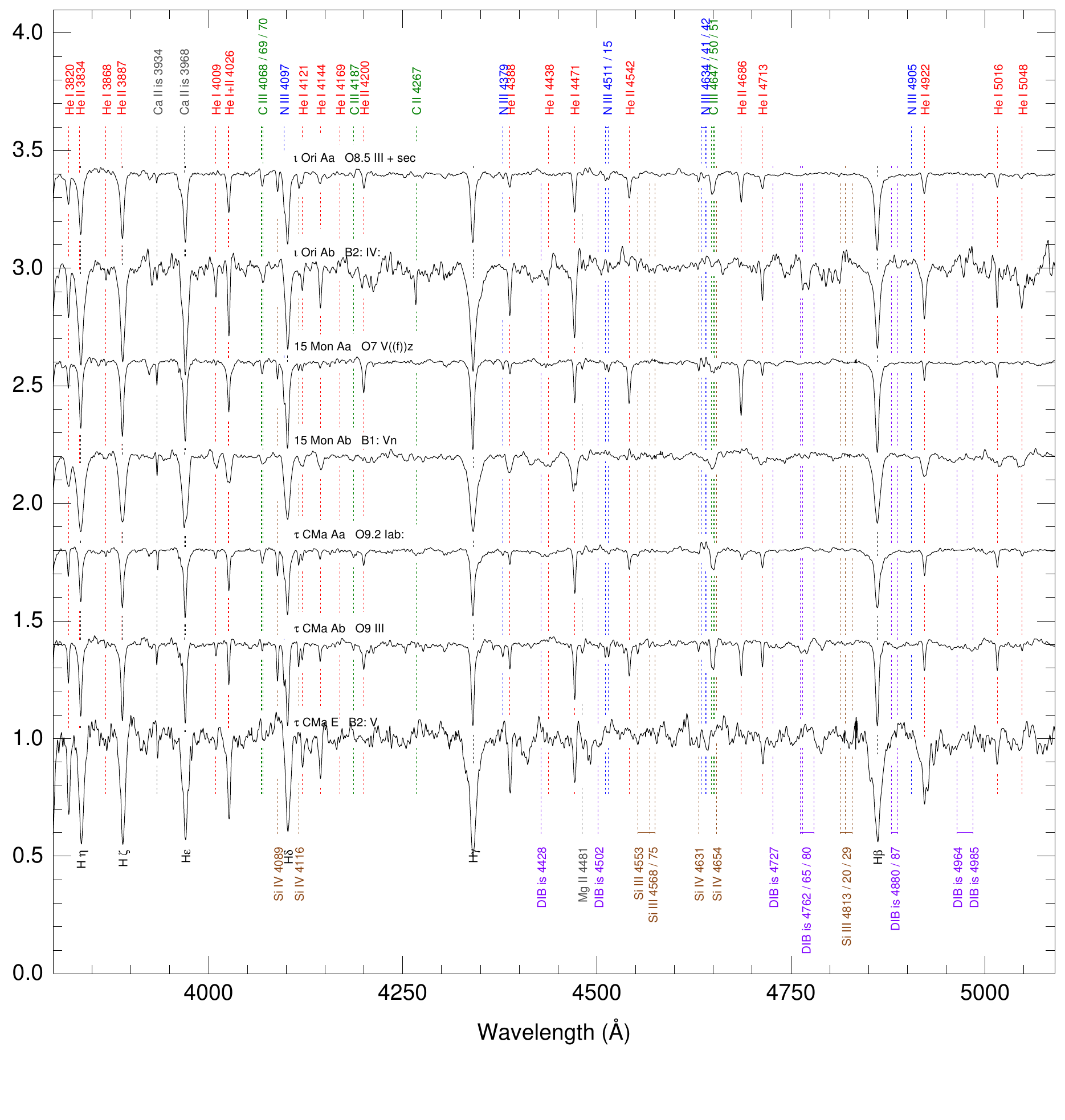}}
\caption{Extracted HST/STIS spectra. The velocity reference system is that of the star in each case and the spectra are plotted at a spectral resolution of 2500.}
\label{spectra}
\end{figure*}

\addtocounter{figure}{-1}

\begin{figure*}
\centerline{\includegraphics[width=\linewidth]{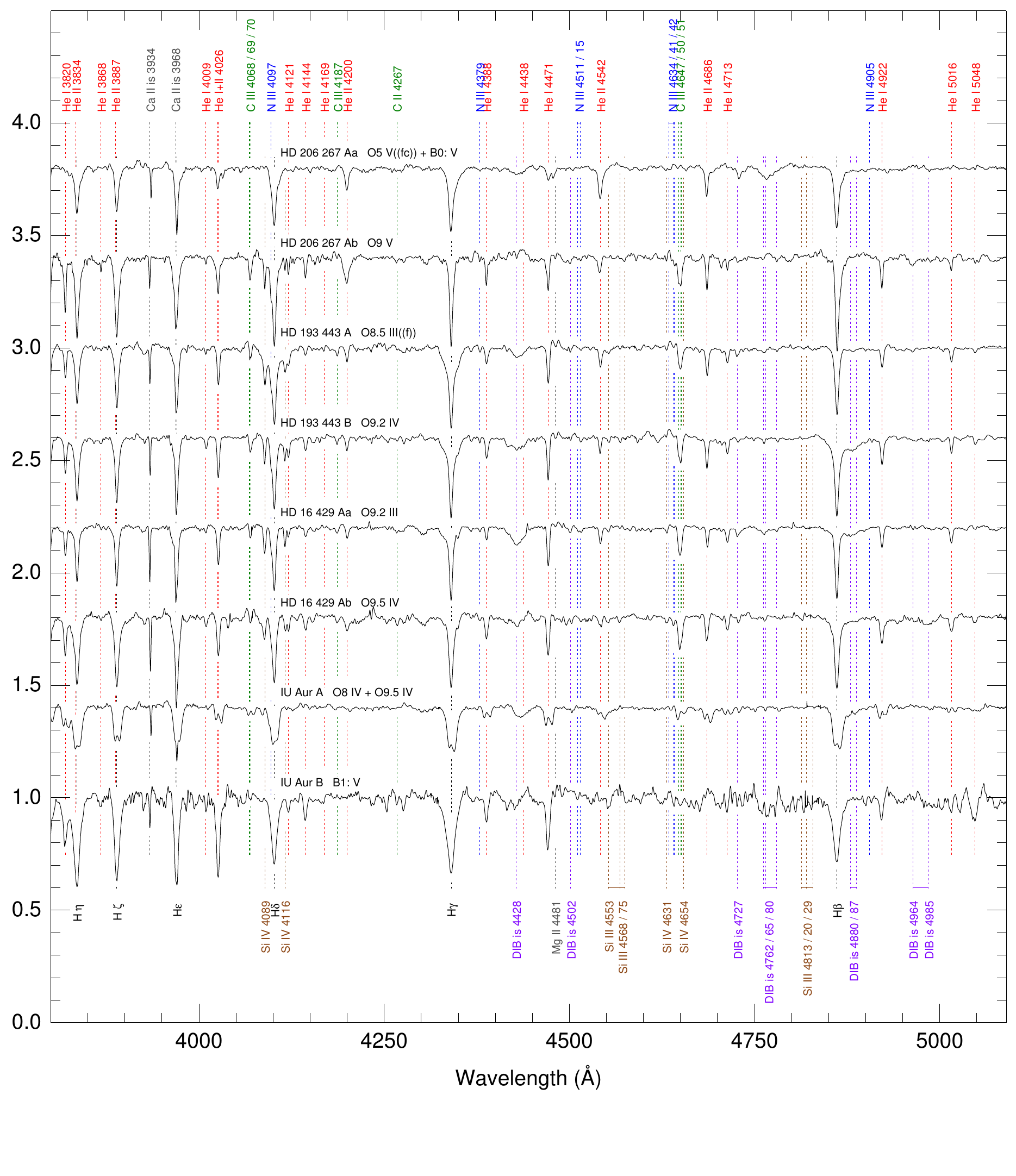}}
\caption{(Continued).}
\end{figure*}

\begin{figure*}
\centerline{\includegraphics[width=\linewidth]{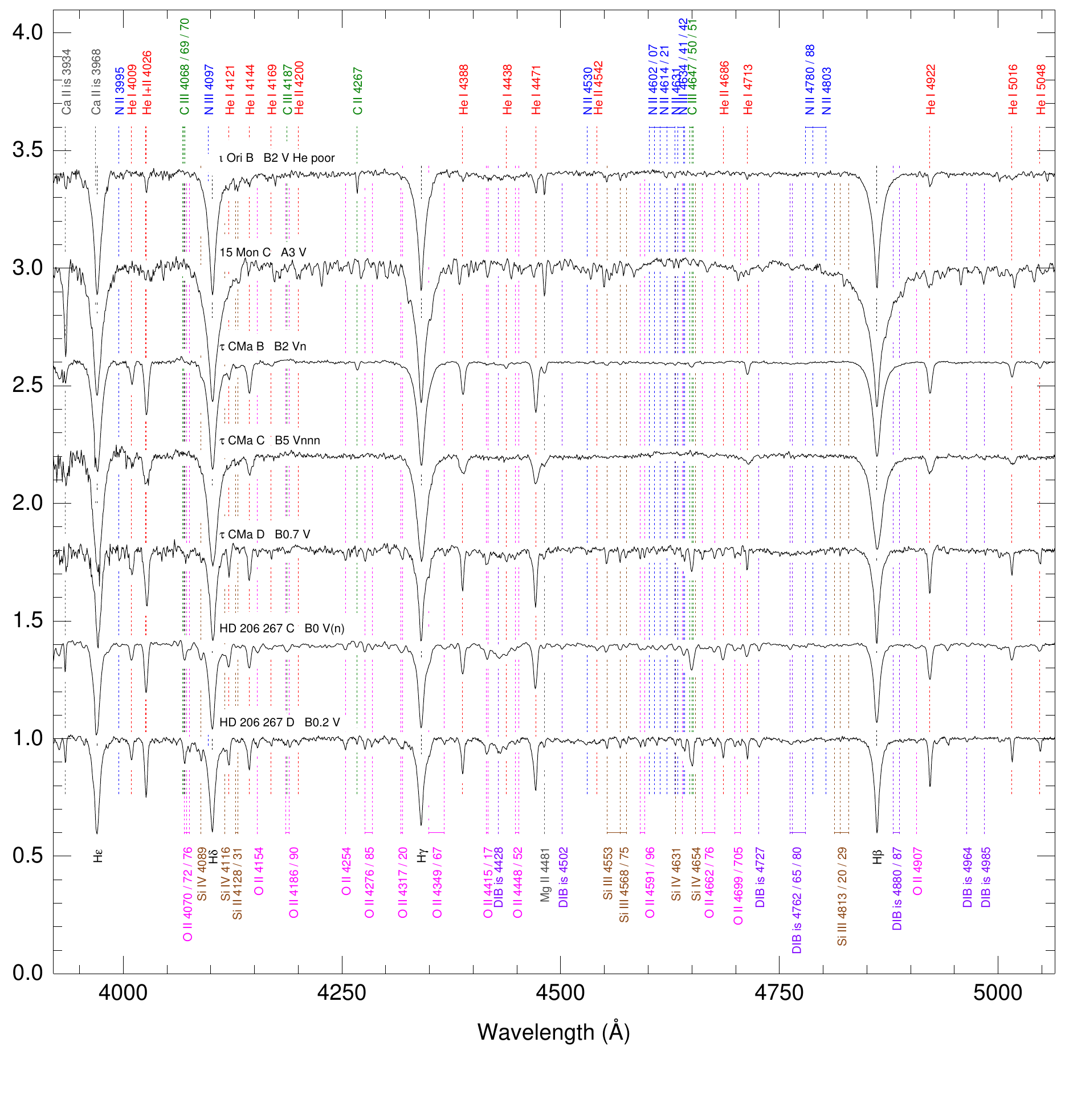}}
\caption{GOSSS spectra.}
\label{spectraGOSSS}
\end{figure*}

\subsection{$\iota$~Ori~Aa,Ab}

$\,\!$\indent This system is composed of at least four stars. Two are in an eccentric 29.1~d period \citep{Marcetal00}, form the visual component Aa, and were classified 
as O8.5~III~+~B0.2~V in MONOS-I. The previously spectroscopically unresolved Ab component is located 155~mas away. The more distant B component is 11\farcs3 away and is a
He-poor B star \citep{ContLoon70}.

In our spectral extraction we fix $d^\prime$ between Aa and Ab to 155~mas
(an estimate from the existing WDS values, which show the separation increasing between the mid 1990s and the last 2016 measurement),
as the secondary is too weak to have its position in the spatial profile fitted 
simultaneously with the profile shape. The $\Delta B$ is the second largest in the HST/STIS sample after $\tau$~CMa~Aa,E.

Some lines in Aa are slightly asymmetric but not to a sufficient degree as to allow us to give a spectral classification to the secondary spectroscopic component.
The primary is clearly O8.5~III, as in MONOS-I. A second epoch may be able to provide a spectral classification for the secondary if the velocity separation is more favorable.

The Ab spectrum is one of the noisiest one in Fig.~\ref{spectra}, as expected from the large $\Delta B$, but is clearly distinct from that of Aa. There are strong He\,{\sc i} 
lines, no He\,{\sc ii} lines or \CIII{4650} (making it B1 or later), and \CII{4267} is strong (while absent in Aa), which is typical around B2. The Balmer lines are somewhat narrower 
than those of a dwarf at that subtype, hence the B2:~IV: classification.

We also analyze here the GOSSS spectrum of the B component, a star with a very low $v\sin i$ and a number of corresponding narrow lines, which is relatively unusual among mid-B
dwarfs, whose spectrum usually has few features. The simultaneous detection of the \SiIId{4128+31} doublet and the \SiIIIt{4553+68+75} triplet allows for a precise determination
of the spectral subtype as B2 independently of any composition anomalies. The strength of \CII{4267}, \MgII{4481}, and several weak N\,{\sc ii} lines confirm the spectral subtype. At
the same time, the Balmer line profiles correspond to a dwarf luminosity class for a B2 star. The big discrepancy lies in the He\,{\sc i} lines, which are very weak when at B2 they 
should be very strong. Hence the He poor qualifier already noted half a century ago \citep{ContLoon70}. 
Note that Simbad lists a variety of spectral types for $\iota$~Ori~B, the reason being that if 
one pays attention just to \HeI{4471}, \MgII{4481}, and the Balmer lines we could easily classify it as B8~III. A cautionary tale here is that if the S/N were lower or $v\sin i$ 
were larger we would easily miss most of the weak lines and think this object is a late-B star.

Summarizing the results from different sources, $\iota$~Ori is a hierarchical system \citep{Toko97}
composed of an O8.5~III~+~B0.2~V spectroscopic binary (Aa) with a mid-B star orbiting 
around them (Ab, classified here for the first time) and another one much farther away (B). The A component has an entry in {\it Gaia}~DR2 but without a parallax. The B 
component, on the other hand, has a {\it Gaia}~DR2 measurement of $\varpi = 2.3839\pm 0.0810$~mas and a good-quality RUWE (renormalized unit weight error, \citealt{Lindetal18b}) 
of 1.07, which allows for a high-quality distance measurement. Using the self-gravitating isothermal Galactic vertical distribution prior of \citet{Maiz01a,Maiz05c} with the updated 
parameters of \citet{Maizetal08a} and correcting for a parallax zero point of $40\pm 10$~$\mu$as, we obtain a distance of $412^{+14}_{-13}$~pc. 
Therefore the three orbits have estimated semi-major axes of 0.38~AU (from the SB2 data and
estimated masses), 70~AU, and 4600~AU (from $d$), respectively, noting the last two values can be significantly different if the orbits have 
orientations far from the plane of sky or high eccentricities. From the last two semi-major axes the periods are estimated around 87~a and 43~ka. The WDS
lists a significant motion of Ab with respect to Aa over a period of 22~a with a $d$ that is currently increasing (something that 
agrees with the value measured here), suggesting a period of several centuries, so the orbit of Ab with respect to Aa appears to be inclined, eccentric, or both. 


\subsection{15~Mon~Aa,Ab}
\label{15Mon}

$\,\!$\indent This object is the ionizing source of the \HII\ region NGC~2264 and is composed of an inner Aa,Ab pair with a period of 108$\pm$12~a \citep{Maiz19} and a
third B component at a distance of 3\farcs0. In \citet{Maizetal18a} we used lucky spectroscopy to assign separate spectral types of O7~V((f))z~var to Aa,Ab and 
B2:~Vn to B but we were not able to separate Aa and Ab when we aligned the slit along them. The \citet{Maiz19} orbit indicates that the system passed through 
periastron in early 1996, achieved a minimum $d$ of 19-20~mas shortly afterwards, and that the separation has increased since then. The WDS lists a large number of additional 
components, many of them likely independent stars in the NGC~2264 cluster, of which the only one within 30\arcsec\ of AaAb is C.

The predicted $d^\prime$ at the time of the HST observation from the \citet{Maiz19} orbit was 135~mas but our fitted value is somewhat larger, 143~mas, suggesting that 
the latest ground-based observations may slightly underestimate it.

We classify Aa as O7~V((f))z as we already did in MONOS-I but dropping the var suffix as we currently only have a single HST epoch. The absorption lines are especially narrow,
indicating a low $v\sin i$.  The spectrum of the Ab component is generally well separated from that of Aa but a small contamination (or rather oversubtraction)
is seen at the bottom of the \HeI{4471}
absorption line. An accurate classification cannot be attained because the object is a fast rotator but based on the absence of both He\,{\sc ii} lines and \CII{4267}, the
presence of weak \SiIII{4553}, and the width of the Balmer lines we arrive at B1:~Vn.

We also classify here the GOSSS spectrum of the weak C component. It is an A3~V star.

15~Mon is another hierarchical system composed of a central O star and two fast rotators of B spectral type. In \citet{Maiz19} we derived a {\it Gaia}~DR2 distance of 
$719\pm 16$~pc, which places the current values of $d$ as 103~AU and 2200~AU for Aa,Ab and Aa,B, respectively. The measured semi-major axis of the 
\citet{Maiz19} Aa,Ab orbit is 112.5~mas (smaller than the current $d$, note that the orbit is quite eccentric), which corresponds to 81 AU.


\subsection{$\tau$~CMa~Aa,Ab,E}

$\,\!$\indent $\tau$~CMa is our southernmost target and, for that reason, was not included in MONOS-I. This is a complex system composed of an inner spectroscopic binary
with a 154.92~d period \citep{StruPogo28,StruKraf54,Sticetal98} orbited by a visual companion for which the latest $d$ values in the WDS are in the 110-120~mas range. In addition, 
one of the components shows eclipses with a much shorter period of 1.282\,122~d \citep{vanLvanG97}. Another 
dim component, E, is listed as 0\farcs94 away in the WDS but note that the current position angles for both the Aa,Ab and Aa,E pairs are offset by 180 degrees in the WDS, 
i.e., Ab is actually to the NW of Aa and E is to the W of Aa, as already pointed out by \citet{Aldoetal15} and confirmed independently in unpublished AstraLux imaging 
(see \citealt{Maiz10a}) and in a short STIS imaging exposure obtained in the same orbit as our spectra. The WDS also lists another three additional components with larger 
$d$: B, C, and D. GOSSS gives a combined spectral classification of O9~II for Aa,Ab \citep{Sotaetal14} which is consistent with the previous result from \citet{Walb72}.

In our spectroscopic data we measure $d^\prime = 95$~mas for Aa,Ab, which is somewhat smaller than the previous values in the WDS. Nevertheless, those other values were 
obtained 6-10~years earlier and older ones are even larger (in the 1960s $d$ appears to have been around 200~mas), so it appears that the system has a 
significant inwards motion in the plane of the sky confirmed by our measurement. The separation along the slit we measure for Aa,E is 685~mas which, corrected for the 
43$^{\rm o}$ misalignment deduced from the imaging data, yields a $d = 937$~mas, consistent with the WDS value.

The most important result for $\tau$~CMa is that both the Aa and Ab components have O-type spectral classifications, something that is shown here for the first time, as the 
\citet{Sticetal98} analysis was for just an SB1 orbit. We classify the weaker Ab component as O9~III. The brighter Aa component is an O9.2 with an uncertain luminosity
classification, as the He/He and Si/He criteria give different results, which is often the case when the spectrum is a composite of a late O and an early B star (e.g. compare 
the $\sigma$~Ori case in \citealt{Sotaetal14} and in MONOS-I). Therefore, the spectroscopic binary is likely to be Aa but we require a second HST epoch to confirm this.

The spectrum of the E component is very noisy due to the large $\Delta B$ or, equivalently, to the short exposure time for its magnitude, as the STIS spatial profile yields
little contamination from Aa,Ab at such large separations. We are only able to derive an uncertain B2:~V classification.

We also analyze here the GOSSS spectrum of the D, B, and C components. They are three B dwarfs of progressively later subtypes: B0.7, B2, and B5, respectively, with their magnitudes 
increasing accordingly. They also follow a progression in $v\sin i$ from slow, to fast, and to very fast values. Indeed, $\tau$~CMa~C has an nnn rotation index, which corresponds 
to a $v\sin i$ close to 500~km/s. Surprisingly for such a fast B-type rotator, there is no sign of emission in any Balmer line (including H$\alpha$, observed but not shown here) so 
it does not appear to be a Be (at least when we observed it, Be emission lines can notoriously appear and disappear).

$\tau$~CMa is a high-order multiple system which is the brightest object at the core of the NGC~2362 cluster. As in other similar cases (e.g. HD~93\,129 in Trumpler~14), it is
difficult to say where the multiple system ends and the cluster begins and only long-term monitoring may provide an answer to that question\ldots or it may not, as such 
circumstances may lead to unstable arrangements with exchanges of partners and/or ejections during the lifetime of the cluster, in which case the question may be ill posed. The
{\it Gaia}~DR2 RUWE for $\tau$~CMa~Aa,Ab is 33.6, so we cannot use its parallax (which is negative, anyway) to estimate its distance. Instead, we use a simplified version of 
the algorithm of \citet{Maiz19} to select the eight brightest stars in NGC~2362 (excluding $\tau$~CMa~Aa,Ab) with similar proper motions and obtain a cluster parallax of 
$0.6265\pm 0.0477$~mas. Applying the same parallax zero point correction and prior as before, we obtain a distance of $1.50^{+0.12}_{-0.10}$~kpc to the cluster.


\subsection{HD~206\,267~Aa,Ab}

$\,\!$\indent This is another complex system. The brightest component, Aa, is an SB2 with a 3.71~d period \citep{Raucetal18}. The Ab and B components were measured to be 
100~mas and 1\farcs8 away, respectively, in MONOS-I but B is too faint to make a significant contribution to the spectrum\footnote{As of the time of this writing, Simbad
claims that B is an O star, but this is likely a confusion with Ab.}. Two other bright components, C and D, are 
farther away. In MONOS-I we give a spectral classification of O6~V(n)((f))~+~B0:~V for Aa,Ab and note that at least one of the spectral types must be a composite given 
the presence of three early-type stars in the aperture.

In our spectroscopic data we measure $d^\prime = 64$~mas and $\Delta B = 0.95$ for Aa,Ab. The separation is significantly smaller than the previous AstraLux 
\citep{Maizetal19b} and WDS values and the short STIS imaging exposures we took in the same orbit as the spectroscopy show the two stars aligned on the slit within 
$\sim 10^{\rm o}$, so a large position angle difference can be discarded. The magnitude difference is also smaller than the one measured with AstraLux but closer to the 1.1-1.3~mag
values of the most recent WDS data. We have revised the \citet{Maizetal19b} data and we think that it is at fault due to the lack of adequate PSF stars (the problem does not
affect the rest of the objects in that paper). Therefore, the separations and magnitude differences for HD~206\,267~Aa,Ab in that paper should not be used but the position 
angles are more likely to be reliable, as those depend less on the PSF choice. In any case, the value of $d^\prime = 64$~mas is clearly smaller than the previous HST value of
97.1~mas (a lower limit) from a 2008 FGS measurement by \citet{Aldoetal15}, indicating that there is a significant motion in the 11~a between the two epochs and that the
period may be of the order of one or two centuries, as already suggested by \citet{Maizetal19b} from the change in position angle.

Our separate spectra confirms that the SB2 is indeed Aa, which was caught at an orbital point with a large velocity difference that allowed the two spectroscopic components
to be classified. The system is O5~V((fc))~+~B0:~V. The primary of the spectroscopic binary is one spectral subtype earlier than the MONOS-I result (which was likely blended
with Ab) and half a spectral subtype than the \citet{Raucetal18} result.

The Ab component, on the other hand, is classified as O9~V. This is the first time this object is directly classified, as previous studies had just guessed its nature.

We also analyze here the GOSSS spectra of the C and D components. Both are very early B-type dwarfs, B0 in the case of C and B0.2 in the case of D. The first one is a moderately
fast rotator and the second one is a slow rotator.

HD~206\,267~Aa,Ab is another high-order multiple system at the center of a cluster, in this case Trumpler~37. The {\it Gaia}~DR2 RUWE for HD~206\,267~Aa,Ab is 2.8, so 
we cannot use its parallax (which has a relatively large uncertainty, anyway) to estimate its distance. Instead, we use the parallaxes of the C and D components to 
obtain a combined value of $0.8807\pm 0.0516$~mas, where the uncertainty includes the covariance term. 
Applying the same parallax zero point correction and prior as before, we obtain a distance of $1093^{+67}_{-59}$~pc to HD~206\,267.


\subsection{HD~193\,443~A,B} 

$\,\!$\indent This object is a triple system composed of a spectroscopic binary with a 7.467~d period \citep{Mahyetal13} and a visual companion located
137.7~mas away \citep{Aldoetal15}. HD~193\,443~A,B is a poorly known system, in part because of the confusion induced by the small $\Delta m$ between the two visual
components A and B. In contrast with previous results, \citet{Aldoetal15} found out a magnitude difference of $-$0.264 i.e. the A (currently eastern) component is dimmer than the
B (currently western) component. \citet{Mahyetal13} classified this system as O9~III/I~+~O9.5~V/III but failed to notice the presence of a third light in the system, so it is not 
clear from their results whether the spectroscopic binary is A or B. \citet{Sotaetal11a,Sotaetal14} could only provide a combined spectral classification of O9~III for
A,B.

As the WDS data suggests that the A,B pair is moving in a clockwise direction (position angles of 290-300$^{\rm o}$ in the early 20th century to 258.5$^{\rm o}$ in 
the late 2008 measurement of \citealt{Aldoetal15}), we set the position angle of our slit at 250$^{\rm o}$. We obtain a $d^\prime = 138$~mas and $\Delta B = -0.26$~mag, which are 
nearly identical to the \citet{Aldoetal15} results.

The derived spectral classifications for A and B are O8.5~III((f)) and O9.2~IV, respectively. No double lines are seen in either spectra but this was expected, as the maximum
velocity separation for the eccentric spectroscopic orbit is not large and the system was caught near the quadrature that takes place far from periastron. We plan to observe it 
again at a more favorable epoch close to the other quadrature. Nevertheless, at the original spectral resolution some of the Ab lines are asymmetric, indicating that is the 
spectroscopic binary. If that were indeed the case, A would be the more massive, luminous, and evolved component but would be dimmer than B due to the existence of two 
stars there. What we can discard is that any of the stars is a supergiant, one of the possibilities mentioned by \citet{Mahyetal13}.

Somewhat differently from some of the previous cases, there are few objects associated with this system. The WDS lists a weak C component 9\arcsec\ away with a {\it Gaia}~DR2
proper motion consistent with being associated with A,B. There is no surrounding cluster, though, and some of the brighter nearby stars have differing proper motions. This
poses a problem for calculating the distance to HD~193\,443, as the {\it Gaia}~DR2 entries for A,B and C have RUWE values of 14.2 and 2.8, respectively, making their
parallaxes useless for a reliable determination. For this object we will have to wait at least until future {\it Gaia} data releases.

\subsection{HD~16\,429~Aa,Ab}

$\,\!$\indent HD~16\,429 is another complex system with an inner pair Aa,Ab separated by 0\farcs28 in MONOS-I and with a significant inwards relative motion. Farther
away we find B, a foreground interloper with an F spectral type, and C, classified as B0.7~V(n) (see MONOS-I). Ab was measured to be fainter than Aa by 2.1-2.2~mag
in MONOS-I. \citet{McSw03} detected the existence of a spectroscopic binary in Aa,Ab and gave the three spectral types 
O9.5~II~+~O8~III-V~+~B0~V?, assigning the first one to Aa and the other two to Ab. 
In MONOS-I we pointed out that such a configuration was difficult to reconcile with the relatively large magnitude difference. 
The GOSSS data of \citet{Sotaetal14} was insufficient to resolve any components and provided only a combined spectral classification of O9~II-III(n)~Nwk for Aa,Ab.

In the STIS spectra we measure $\Delta B = 2.02$~mag and $d^\prime = 275$~mas for the Aa,Ab pair, values that are compatible with the MONOS-I results and confirm 
that the two components are approaching each other.

We obtain a spectral classification of O9.2~III for Aa and of O9.5~IV for Ab. We do not see double or even asymmetric lines in either object, indicating we caught it far from
either quadrature, which is consistent with the orbital phase predicted by the \citet{McSw03} analysis. 
The first classification is relatively similar to the \cite{McSw03} but the second one is considerably later than
her result. One possibility would be that indeed Ab is the spectroscopic binary and our O9.5 classification is just a composite between earlier- and later-type stars.
There are two issues with that. The first one is the previously mentioned magnitude difference, which makes it hard to accommodate two O/early-B stars in Ab and just one in Aa. The
second one are the velocities. \citet{McSw03} obtains a systemic velocity of $-$50~km/s for HD~16\,429~Aa,Ab and that is close to the value we measure for Ab. On the other
hand, for Aa we obtain a value close to $+$20~km/s. Therefore, the evidence points towards Aa being the spectroscopic binary. To confirm this we plan to obtain additional STIS
epochs of the system. 

The literature lists HD~16\,429 as a multiple system but the {\it Gaia}~DR2 data hints at the existence of a small cluster associated with it, as there are several stars that
appear to follow a young isochrone at the same distance. We cannot use the {\it Gaia}~DR2 data for Aa,Ab itself because, as with most of the systems in this paper, the
unaccounted visual multiplicity in the {\it Gaia} pipeline translates into a large RUWE value. On the other hand we cannot use the data for the B component because it is
a foreground chance alignment. The C component is a more promising candidate and appears to be the brightest member of the cluster with good quality data. As we did for 
$\tau$~CMa, we use a simplified version of the algorithm of \citet{Maiz19} to select the three brightest stars in the vicinity (using C as the reference) with similar proper 
motions and obtain a cluster parallax of $0.4657\pm 0.0480$~mas. Applying the same parallax zero point correction and prior as before, we obtain a distance of 
$2.04^{+0.23}_{-0.19}$~kpc to the cluster.

\subsection{IU~Aur~A,B}

$\,\!$\indent The final target in our sample is also complex. The inner pair A,B is separated by 128-144~mas in the MONOS-I AstraLux data with a trend that suggests that
the two components are approaching each other. Two weak (C and D) components are detected with AstraLux farther away. The A component is an eclipsing binary
with relatively deep and similar primary and secondary eclipses \citep{Ozdeetal03}. MONOS-I measured a variable magnitude difference between A and B of 1.40-2.06~mag, 
with most of the variation likely caused by the eclipses. The period of the eclipsing binary is 1.811~d \citep{Drecetal94} but the eclipses are modulated by an eccentric 
orbit with a 293.3~d period \citep{Ozdeetal03} which must be caused by a dim or dark object distinct from A and B that does not cause a significant signal in the 
observed spectrum (see MONOS-I). In MONOS-I we found that the primary apparently changes its spectral type between O8.5~III(n) and O9.5~IV(n) 
while the secondary also changes (but to a lesser degree) between O9.7~IV(n) and O9.7~V(n)
but we cautioned that those changes may be caused not only by the proximity between the two stars and its associated orbital effects
\citep{Harretal98} but also by the different degrees of contamination by the B component.

The STIS spectra yield a $\Delta B$ of 2.04~mag but we had to fix $d^\prime$ due to the large $\Delta B$. The system was caught near quadrature,
so we may consider that to be the uneclipsed magnitude difference, which agrees with the AstraLux results in MONOS-I.

Being able to eliminate the effect of the B component, we classify A as O8~IV(n)~+~O9.5~IV(n). As we suspected, the B component was making the two stars in the eclipsing binary
appear as having a later spectral type. 

The spectrum of the B component is relatively noisy but it is well separated from that of A, as all the lines have a single kinematic component in clear contrast to the two
seen in A with a separation of $\sim$475~km/s. The derived spectral type is B1:~V.

There are several massive stars (e.g. HD~35\,619) and H\,{\sc ii} regions close to IU~Aur but it is not clear whether our target is associated with any of them. The bad 
quality of the {\it Gaia}~DR2 astrometric fit (RUWE of 9.9) impedes the determination of its distance. As it was the case for HD~193\,443, we will have to wait for future
{\it Gaia} data releases.

\section{Analysis and future work}

\begin{figure}
\centerline{\includegraphics[width=\linewidth]{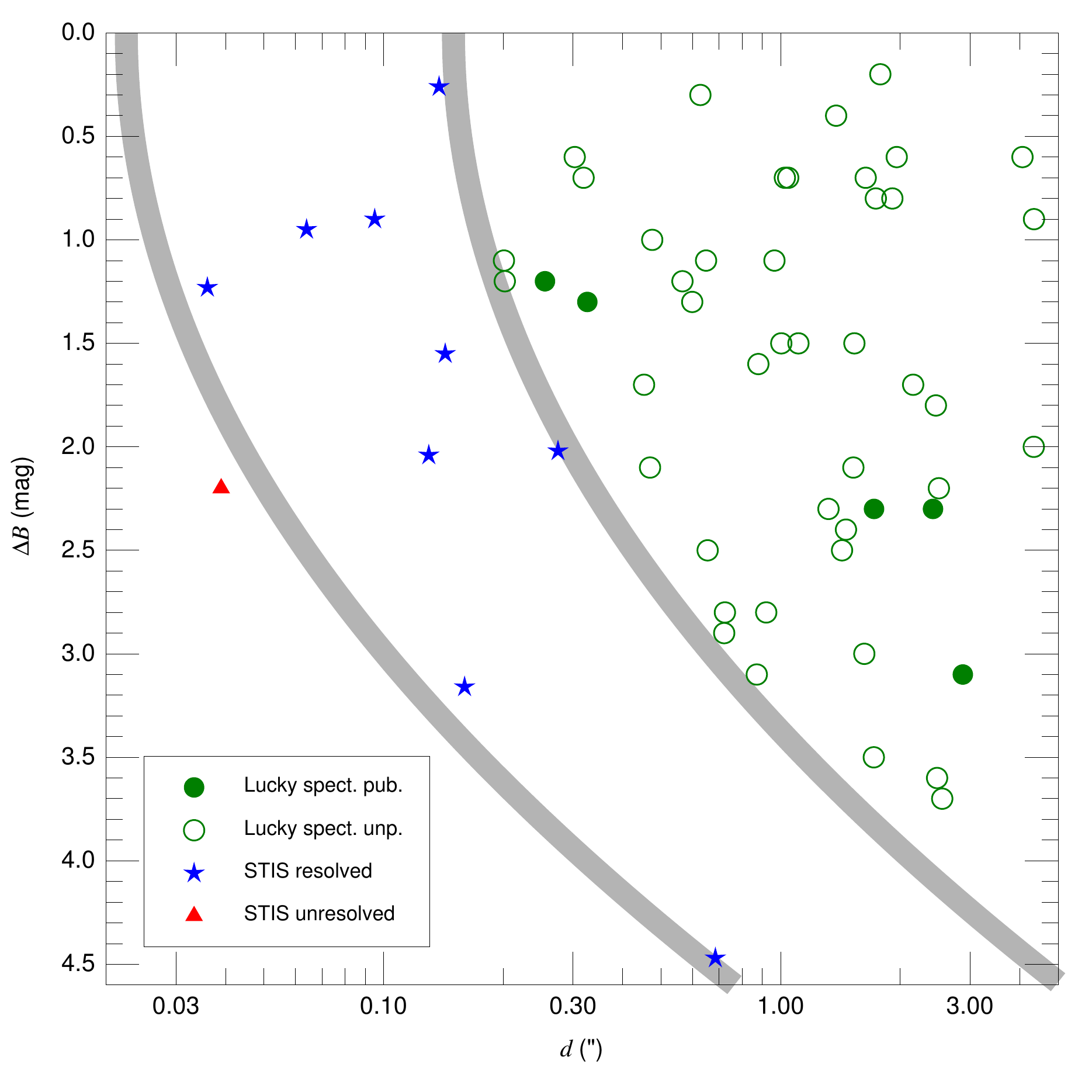}}
\caption{Binary systems resolved using lucky spectroscopy with WHT/ISIS (published in \citealt{Maizetal18a} and unpublished) and HST/STIS and unresolved systems using 
HST/STIS in the $d-\Delta B$ plane. The gray curves mark the tentative empirical boundaries between the regions in the plane accessible with different techniques: the 
left one is for HST/STIS and the right one for lucky spectroscopy with WHT/ISIS.}
\label{sep_DB}
\end{figure}

$\,\!$\indent In this paper we have used HST/STIS to obtain spatially resolved spectroscopy of the close components of seven visual multiple systems containing at least one O star.
We have failed to spatially resolve an eighth system, $\zeta$~Ori~Aa,Ab,
which is located in the most difficult region of the $d-\Delta B$ plane (Fig.~\ref{sep_DB}). Previously, we had similarly
resolved five similar systems using ground-based lucky spectroscopy with the WHT \citep{Maizetal18a}. Since then, we have used the same WHT setup to successfully separate several 
tens of similar systems (Ma{\'\i}z Apell\'aniz et al. in preparation). In Fig.~\ref{sep_DB} we show the location of all those systems in the $d-\Delta B$ plane along with tentative
empirical boundaries for each of the two techniques: HST/STIS and lucky spectroscopy with WHT/ISIS. Even though they are not plotted, we have attempted and failed to separate with
lucky spectroscopy several targets located in the region between the two boundaries (see section~\ref{15Mon} for an example).

As expected, both techniques can reach the smallest separation when the magnitude difference is small, a pattern that is repeated for every technique that attempts to resolve
closely separated point sources and that depends on how narrow and stable the PSF is. Also as expected, HST/STIS provides a significant improvement in separation with respect to
lucky spectroscopy, which we can quantify as a factor of $\approx 8$ (at the same $\Delta B$). This happens because the STIS PSF is narrower and more stable, being adequately
defined by a diffraction pattern with slight modifications from small focus changes. In lucky spectroscopy, the PSF depends on the observing conditions (which are not always optimal)
and even when seeing is exceptional there are wings that hamper the measurement of weak companions. Nevertheless, lucky spectroscopy has some advantages with respect to HST/STIS,
namely the smaller fraction of time spent on overheads, the higher number of photons collected per final unit time (with the corresponding increase in S/N), and being a
cheaper technique that allows for larger samples.

Another lesson of this work is that the {\it Gaia}~DR2 astrometric results of unresolved multiple systems are suspect. This is a well known effect when the companion is a
dark object \citep{Andretal19,SimDetal20} but it is also noticeable when there are two or more moving bright objects blended into a Gaia source, which is the case for a majority of
the O stars in the Galaxy given the limitations of the telescope \citep{BranCata19}. An advice here is that one should not analyze data from the {\it Gaia} archive without reading the
fine print e.g. using the RUWE to estimate the validity of the astrometry. The situation should change in the future when the {\it Gaia} epoch astrometry is released, as one would
be able to fit stars of interest with binary parameters, possibly constrained by external information such as spectroscopic or visual orbits.

We still have twelve remaining visits in HST GO program 15\,815 that will be used to observe two more systems ($\theta^1$~Ori~Ca,Cb and HD~193\,322~Aa,Ab) and obtain new epochs for
some of the targets in this paper. We also plan to request more HST time to extend the sample to other massive stars with close visual companions. Finally, we want to extend the 
lucky spectroscopy technique to other ground-based telescopes. Overall, in the next few years we hope to obtain spatially resolved spectroscopy of $\sim$100 visual binaries and, in
that way, significantly improve our statistics of such systems.

\begin{acknowledgements}
J.M.A. acknowledges support from the Spanish Government Ministerio de Ciencia, Innovaci\'on y Universidades through grant PGC2018-095\,049-B-C22. 
R.H.B. acknowledges support from DIDULS Project 18\,143.
Based on observations made with the ESA/NASA Hubble Space Telescope, obtained from the data archive at the Space Telescope Science Institute. 
STScI is operated by the Association of Universities for Research in Astronomy, Inc. under NASA contract NAS 5-26\,555. 
The GOSSS spectroscopic data in this article were gathered with three facilities: 
the 1.5~m Telescope at the Observatorio de Sierra Nevada (OSN), 
and the 2.0~m Liverpool Telescope (LT) and 4.2~m William Herschel Telescope (WHT) at the Observatorio del Roque de los Muchachos (ORM). 
We thank an anonymous referee for useful suggestions,
Amber Armstrong and the rest of the STScI staff for their help in implementing a technically demanding program, Brian D. Mason for providing us with the WDS data, 
and Alfredo Sota for the processing of the OSN and WHT data.
\end{acknowledgements}

\vfill

\eject

\bibliographystyle{aa} 
\bibliography{general} 

\end{document}